\documentclass[sigconf]{acmart}
\usepackage{booktabs} 

\usepackage[english]{babel}
\usepackage{moresize}
\usepackage{amsmath}
\usepackage{algorithmic}
\usepackage{balance}
\usepackage{comment}
\usepackage{paralist}
\usepackage{bm}
\usepackage{pgfplots}
\usetikzlibrary{pgfplots.dateplot}

\usepackage{fancyhdr}
\usepackage{flushend}
\usepackage[english]{babel}
\usepackage[latin1]{inputenc}
\usepackage{mathrsfs}
\usepackage{graphicx}

\usepackage{amssymb}
\usepackage{amsfonts}
\usepackage{url}
\usepackage{longtable}
\usepackage{rotating}
\usepackage{multirow}
\usepackage{mathrsfs}
\usepackage{subfigure}
\usepackage{enumitem}
\usepackage[linesnumbered,algoruled,boxed,lined]{algorithm2e}
\usepackage{adjustbox}
\usepackage{hyperref}
\usepackage{pgfplots}
\usetikzlibrary{pgfplots.dateplot}
\usepackage{filecontents}
\definecolor{tblue}{RGB}{31,119,180}
\definecolor{torange}{RGB}{255,127,14}
\definecolor{tgreen}{RGB}{44,160,44}
\definecolor{tred}{RGB}{214,39,40}
\definecolor{tpurple}{RGB}{148,103,189}

\newcommand{\hide}[1]{} 

\newcommand{\ie}{\textit{i}.\textit{e}.}
\newcommand{\eg}{\textit{e}.\textit{g}.} 
\newcommand{\wrt}{\textit{w}.\textit{r}.\textit{t}}

\def\model{DCCF}

\copyrightyear{2023}
\acmYear{2023}
\setcopyright{acmlicensed}\acmConference[SIGIR'23]{Proceedings of the 46th International ACM SIGIR Conference on Research and Development in Information Retrieval}{July 23--27, 2023}{Taipei, Taiwan}
\acmBooktitle{Proceedings of the 46th International ACM SIGIR Conference on Research and Development in Information Retrieval (SIGIR'23), July 23--27, 2023, Taipei, Taiwan}
\acmPrice{15.00}
\acmDOI{10.1145/3539618.3591665}
\acmISBN{978-1-4503-9408-6/23/07}

\begin{document}



\title{Disentangled Contrastive Collaborative Filtering}

\author{Xubin Ren}
\affiliation{%
  \institution{University of Hong Kong}
  \city{Hong Kong}
  \country{China}
  }
\email{xubinrencs@gmail.com}

\author{Lianghao Xia}
\affiliation{%
  \institution{University of Hong Kong}
  \city{Hong Kong}
  \country{China}}
\email{aka\_xia@foxmail.com}

\author{Jiashu Zhao}
\affiliation{%
  \institution{Wilfrid Laurier University}
  \city{Waterloo}
  \country{Canada}}
\email{jzhao@wlu.ca}

\author{Dawei Yin}
\affiliation{%
  \institution{Baidu Inc}
  \city{Beijing}
  \country{China}}
\email{yindawei@acm.org}

\author{Chao Huang}
\authornote{Chao Huang is the corresponding author.}
\affiliation{%
  \institution{University of Hong Kong}
  \city{Hong Kong}
  \country{China}}
\email{chaohuang75@gmail.com}

\renewcommand{\shortauthors}{Xubin Ren, Lianghao Xia, Jiashu Zhao, Dawei Yin, \& Chao Huang}



\begin{abstract}
Recent studies show that graph neural networks (GNNs) are prevalent to model high-order relationships for collaborative filtering (CF). Towards this research line, graph contrastive learning (GCL) has exhibited powerful performance in addressing the supervision label shortage issue by learning augmented user and item representations. While many of them show their effectiveness, two key questions still remain unexplored: i) Most existing GCL-based CF models are still limited by ignoring the fact that user-item interaction behaviors are often driven by diverse latent intent factors (\eg, shopping for family party, preferred color or brand of products); ii) Their introduced non-adaptive augmentation techniques are vulnerable to noisy information, which raises concerns about the model's robustness and the risk of incorporating misleading self-supervised signals. In light of these limitations, we propose a \underline{D}isentangled \underline{C}ontrastive \underline{C}ollaborative \underline{F}iltering framework (\model) to realize intent disentanglement with self-supervised augmentation in an adaptive fashion. With the learned disentangled representations with global context, our \model\ is able to not only distill finer-grained latent factors from the entangled self-supervision signals but also alleviate the augmentation-induced noise. Finally, the cross-view contrastive learning task is introduced to enable adaptive augmentation with our parameterized interaction mask generator. Experiments on various public datasets demonstrate the superiority of our method compared to existing solutions. Our model implementation is released at the link \color{blue}{\url{https://github.com/HKUDS/DCCF}}.
\end{abstract}

\maketitle

\section{Introduction}
\label{sec:intro}

Recommender systems have become fundamental services for suggesting personalized items to users by learning their preference from historical interactions~\cite{wu2020graph,chang2021sequential}. Graph neural networks have recently achieved remarkable success in collaborative filtering (CF) modeling user-item interaction with high-order connectivity, such as NGCF~\cite{wang2019neural},MCCF~\cite{wang2020multi}, LightGCN~\cite{he2020lightgcn}, and GCCF~\cite{chen2020revisiting}. Those GNN-based CF models encode user-item bipartite graph structures into representations via iterative message passing for collaborative information aggregation~\cite{wang2022profiling}. By capturing the high-order user (item) similarity in latent embedding space, graph neural CF methods have provided state-of-the-art recommendation performance.

However, user-item interactions, which serve as important labels for supervised recommendation models, are often highly sparse in real-world recommender systems~\cite{zhou2020s3,xia2023automated,zhu2022mutually}. To address the issue of supervision shortage in recommendations, recent works~\cite{wu2021self,xia2022hypergraph} attempt to marry the power of contrastive learning with GNNs to explore the unlabeled information and offer self-supervision signals. These graph contrastive learning (GCL) methods propose to learn invariant user (item) representations by maximizing agreement between established contrastive augmentation views. In general, by following the mutual information maximization principle~\cite{velickovic2019deep,peng2020graph}, the agreements of pre-defined positive pairs are achieved, and embeddings of negative pairs are pushed apart. Two key research lines of augmentation schemes have recently emerged in GCL-based collaborative filtering. To be specific, SGL~\cite{wu2021self} generates contrastive views with stochastic augmentors, \eg, random node/edge dropout. To supplement the direct graph connections, HCCF~\cite{xia2022hypergraph} and MHCN~\cite{yu2021self} propose to purse the consistency between node-level representations and graph-level semantic embeddings.

Although promising results have been achieved, we argue that two key limitations exist in current GCL recommender systems. \\\vspace{-0.12in}

\emph{First}, most previous studies have ignored the fact that the latent factors behind user-item interactions are highly entangled due to preference diversity, resulting in suboptimal augmentation-induced user representations. In real-life applications, the formation of user-item interactions is driven by many intent factors~\cite{wang2020disentangled,wang2020disenhan}, such as purchasing products for a family party or being attracted to certain clothing characteristics. However, the learned user preferences with the encoded invariant representations in current GCL-based recommendation approaches are entangled, making it difficult to capture the finer-grained interaction patterns between users and items. This hinders the recommender's ability to capture genuine user preferences and provide accurate intent-aware self-supervision. Therefore, there is an urgent need for a new method that can generate disentangled contrastive signals for informative augmentation.\\\vspace{-0.12in}

\emph{Second}, many existing GCL-based methods still struggle to provide accurate self-supervised learning (SSL) signals against data noise, which makes it difficult to adapt contrastive learning to user-item interaction graphs with diverse structures. Specifically, the introduced stochastic augmentation strategy~\cite{wu2021self} may not preserve the original semantic relationships well, as they use random dropout operators. For example, For instance, dropping hub nodes can damage important inter-community connection structures, resulting in an augmented user-item relation graph that may not be positively related to the original interaction structures. Additionally, Additionally, although some methods incorporate graph-level semantics into auxiliary self-supervised signals~\cite{xia2022hypergraph,yu2021self} via self-discrimination over all nodes, their model performance is vulnerable to user interaction data noise, such as misclicks or popularity bias. Under a contrastive augmentation framework, if the importance of node- or edge-wise SSL signals is not differentiated, methods can be easily biased by supplementing the main recommendation task with self-supervised signals derived from noisy nodes, \eg, users with many misclick behaviors or high conformity to popularity bias~\cite{wang2021deconfounded}.

In this paper, we propose a new disentangled contrastive learning-based collaborative filtering model, called \model, to address the limitations of existing methods. Specifically, Our model encodes multi-intent representations by considering the global dependencies between users and items. We achieve this by designing intent-aware information passing and aggregation between patch-level nodes and global-level intent prototypes. We aim to identify important graph structural information that captures accurate and helpful environment-invariant patterns with intent disentanglement. In this way, we can prevent the distillation of self-supervised information with severe noisy signals. To achieve our goal, we create parameterized edge mask generators that capture implicit relationships among users and items, and we inject intent-aware global dependencies. As a result, the graph structure masker can naturally capture the importance of each interaction for contrastive augmentation, which is adaptive to the user-item relations.

To sum up, the main contributions of this work are as follows:

\begin{itemize}[leftmargin=*]

\item In this work, we study the generalization problem of GCL-based recommender systems in a more challenging yet practical scenario: adapting graph contrastive learning to intent disentanglement with self-supervision noise for collaborative filtering.\\ \vspace{-0.12in}

\item We develop a new recommendation model called \model, with parameterized mask generators that are adaptive to build over the global context-enhanced disentangled GNN architecture. This enhances recommender robustness and generalization ability.\\ \vspace{-0.12in}

\item Extensive experimental results demonstrate that our new method achieves superior recommendation performance compared to more than 10 existing solutions. Furthermore, the effectiveness of our disentangled adaptive augmentation is justified by studies of model ablation, robustness, and interpretability. \vspace{-0.12in}

\end{itemize}

\section{Related Work}
\label{sec:relate}

\noindent \textbf{GNNs-based Recommender Systems}.
Graph neural networks (GNNs) have demonstrated strong performance in representation learning of user preference for recommendation. These GNN-based recommenders perform recursive message passing over graph structures to model high-order collaborative relations~\cite{wu2020graph,yang2021enhanced,chen2023heterogeneous}. Towards this line, Many efforts have been made to build recommender systems based on various graph neural techniques. For instance,  graph convolutional networks have been widely adopted as encoders to model the user-item interaction graph, such as LightGCN, LR-GCCF~\cite{chen2020revisiting}, and HGCF~\cite{sun2021hgcf}. Additionally, graph-enhanced attention mechanisms explicitly distinguish influence for embedding propagation among neighboring nodes, and serve as important components in various recommenders, including social relation learning DGRec~\cite{song2019session}, multi-behavior recommendation~\cite{yang2022multi}, knowledge graph-based recommenders KGAT~\cite{wang2019kgat}, JNSKR~\cite{chen2020jointly}. \\\vspace{-0.12in}

\noindent \textbf{Recommendation with Disentangled Representations}.
Learning disentangled representations of user latent intents from implicit feedback has been a popular topic in recent years. Various approaches have been proposed, such as using variational auto-encoders to encode high-level user intentions for improving recommendation~\cite{ma2019learning}. DGCF~\cite{wang2020disentangled} builds upon this idea of intent disentanglement, and performs disentangled representation learning over graph neural network with embedding splitting. To incorporate side information from user or item domain into recommendation, DisenHAN~\cite{wang2020disenhan} attempts to learn disentangled user/item representations with heterogeneous graph attention network. KGIN~\cite{wang2021learning} is a method that aims to encode latent user intents using item knowledge graph to improve recommendation performance. DCF~\cite{chen2021curriculum} decomposes users and items into factor-level representations and using a factor-level attention mechanism to capture the underlying intents. In CDR~\cite{chen2021curriculumnips}, a dynamic routing mechanism is designed to characterize correlations among user intentions for embedding denoising. However, most existing disentangled recommender systems are built in a fully supervised manner, which can be limited by the sparsity of user-item interactions in real-world scenarios. To address this challenge, we propose a new model that leverages self-supervised learning for intent-aware augmentation. \\\vspace{-0.12in}

\noindent \textbf{Contrastive Learning in Recommendation}. 
Recently, contrastive learning (CL) has gained considerable attention in various recommendation scenarios, such as sequential recommendation~\cite{chen2022intent}, knowledge graph-enhanced recommendation~\cite{zou2022improving}, multi-interest recommendation~\cite{zhang2022re4} and multi-behavior recommendation~\cite{wei2022contrastive}. The most relevant research line in recommendation systems is to enhance graph neural network (GNN)-based collaborative filtering with contrastive learning. To this end, several recently proposed models, such as SGL~\cite{wu2021self}, NCL~\cite{lin2022improving}, and HCCF~\cite{xia2022hypergraph}, have achieved state-of-the-art performance by leveraging contrastive augmentation. For example, SGL~\cite{wu2021self} uses random dropout operators to corrupt interaction graph structures for augmentation. In NCL~\cite{lin2022improving}, representation alignment is performed between individual users and semantic-centric nodes. While these models have been effective in improving recommendation accuracy, they may fall short in encoding latent factors behind user-item interactions, which can result in suboptimal representations with coarse-grained user preference modeling for recommendation.
\section{Methodology}
\label{sec:solution}

\begin{figure*}
    \centering
    \includegraphics[width=1.0\textwidth]{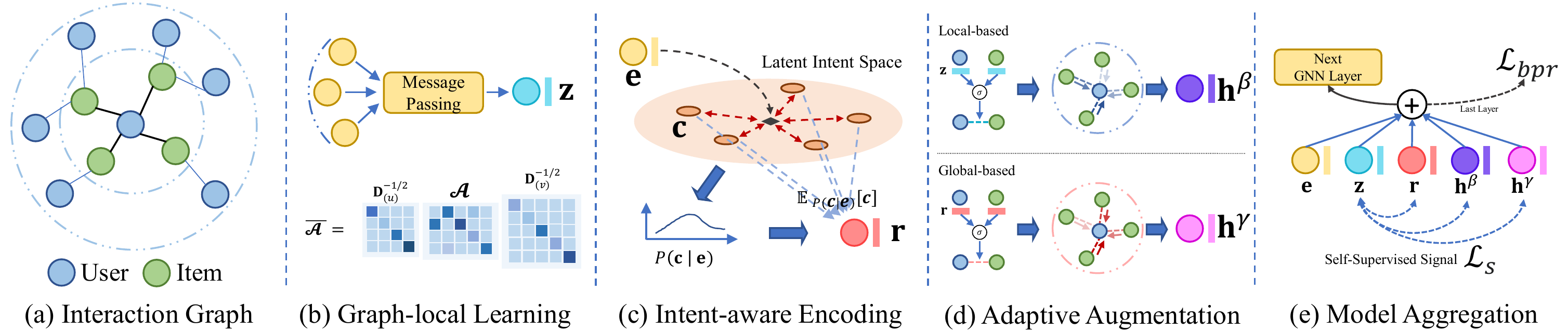}
    \vspace{-0.2in}
    \caption{The overall framework of our proposed \model\ model involves adaptive augmentation through the integration of global intent disentanglement and interaction pattern encoding, resulting in disentangled environment-invariant representations.}
    \vspace{-0.1in}
    \label{fig:framework}
\end{figure*}

\subsection{Disentangled Intent Representation}

\subsubsection{\bf Modeling Latent Intent Factors}
In our recommendation scenario, we represent the interaction matrix between the user set $\mathcal{U}={u_1, ..., u_i, ..., u_I}$ (with size $I$) and item set $\mathcal{I}={v_1, ..., v_j, ..., v_J}$ (with size $J$) as $\mathcal{A} \in \mathbb{R}^{I \times J}$. The entry $\mathcal{A}_{i,j} \in \mathcal{A}$ is set to 1 if user $u_i$ has adopted item $v_j$ before, and $\mathcal{A}_{i,j}=0$ otherwise. Our model aims to predict the likelihood that a candidate user will adopt an item given their observed interactions. From a probabilistic perspective, our predictive model aims to estimate the conditional probability $P(y|u_i, v_j)$ for the interaction between user $u_i$ and item $v_j$, where $y$ is the learned preference score.

When interacting with items, users often have diverse intents, such as preferences for specific brands or interests in the genres and actors of movies~\cite{zhao2022multi,mu2021knowledge}. To capture these diverse intents, we assume $K$ different intents $c_u$ and $c_v$ from the user and item sides, respectively. The intent on the item side can also be understood as the context of the item, for example, a user who intends to shop for Valentine's Day may have a preference for items that have a ``romantic'' context. Our predictive objective of user-item preference can be presented as follows:
\begin{align}\label{eq4}
    \int_{c_{u}}\ \int_{c_{v}}\ P(y, c_{u}, c_{v}|u, v)\, dc_{v}\, dc_{u} = \sum_{k}^{K} P(y, c_{u}^{k}, c_{v}^{k}|u, v)
\end{align}

The user-item interaction probability $y$ is determined by the latent intents $c_u$ and $c_v$ and can be derived using the formulas:
\begin{align}\label{eq7}
    \sum_{k}^{K} P(y, c_{u}^{k}, c_{v}^{k}|u, v) &= \sum_{k}^{K} P(y|c_{u}^{k}, c_{v}^{k}) P(c_{u}^k|u) P(c_{v}^k|v) \\
    &= \mathbb{E}_{P(c_{u}|u)P(c_{v}|v)}[P(y|c_{u}, c_{v})]\label{eq8}.
\end{align}
\noindent Here, we use $f(\cdot)$ to denote the forecasting function over the encoded intents. Following the statistical theory in~\cite{wang2020visual,wang2021deconfounded}, we make the following approximation to derive our prediction objective:
\begin{align}\label{eq9}
    \mathbb{E}_{P(c_{u}|u)P(c_{v}|v)}[f(c_{u}, c_{v})] \approx f(\mathbb{E}_{P(c_{u}|u)}[c_{u}],  \mathbb{E}_{P(c_{v}|v)}[c_{v}]).
\end{align}
\noindent With the above inference, the approximation error, known as \emph{Jensen gap}~\cite{abramovich2016some}, can be well bounded in our forecasting function $f(\cdot)$~\cite{gao2017bounds}.

\subsubsection{\bf Multi-Intent Representation with Global Context}
While intent diversity has been encoded in existing recommender systems through disentangled representations, global-level intent-aware collaborative relations have been largely overlooked. Global-level user (item) dependency modeling can enhance the robustness of GNN-based message passing models against sparsity and over-smoothing issue, via propagating information without the limitation of direct local connections~\cite{xia2022hypergraph}. Towards this end, we propose to disentangle collaborative relations among users and items with both local- and global-level embedding for information propagation. \\\vspace{-0.12in}

\noindent \textbf{Graph-based Message Passing}. 
Owing to the strength of graph neural networks, GNNs has become the prevalent learning paradigm to capture collaborative filtering signals in state-of-the-art recommender systems. Examples include LightGCN~\cite{he2020lightgcn}, LR-GCCF~\cite{chen2020revisiting}, and HGCF~\cite{sun2021hgcf}. The insights offered by these studies have inspired us to build our \model\ model using a graph-based message passing framework for user representations. In general, our message propagation layer is formally presented with the user/item embedding matrix $\textbf{E}^{(u)}\in\mathbb{R}^{I\times d}$ and $\textbf{E}^{(v)} \in \mathbb{R}^{J\times d}$ as follows:

\begin{align}\label{eq1}
    \textbf{Z}^{(u)} = \bar{\mathcal{A}} \cdot \textbf{E}^{(v)}, ~~~
    \textbf{Z}^{(v)} = \bar{\mathcal{A}}^T \cdot \textbf{E}^{(u)},
\end{align}
\noindent The aggregated representations from neighboring nodes to the target ones are denoted by $\textbf{Z}^{(u)} \in\mathbb{R}^{I\times d}$ and $\textbf{Z}^{(v)} \in\mathbb{R}^{J\times d}$. Here, $\bar{\mathcal{A}}\in \mathbb{R}^{I\times J}$ denotes the normalized adjacent matrix which is derived from the user-item interaction matrix $\mathcal{A}$ as $\bar{\mathcal{A}} = \textbf{D}^{-1/2}_{(u)} \cdot \mathcal{A} \cdot \textbf{D}^{-1/2}_{(v)}$.

\noindent where $\textbf{D}_{(u)}\in\mathbb{R}^{I\times I}$ and $\textbf{D}_{(v)}\in\mathbb{R}^{J\times J}$ are diagonal degree matrices.\\\vspace{-0.12in}

To exploit high-order collaborative filtering signals, we perform GNN-based embedding propagation across different graph layers, such as from the $(l-1)$-th to the $(l)$-th layer, as follows:
\begin{align}\label{eq3}
    \textbf{E}_{l}^{(u)} = \textbf{E}_{l-1}^{(u)} + \textbf{Z}_{l-1}^{(u)}, ~~~
    \textbf{E}_{l}^{(v)} = \textbf{E}_{l-1}^{(v)} + \textbf{Z}_{l-1}^{(v)},
\end{align}
\noindent To suppress the over-smoothing effect, residual connections are applied to the aggregation phase~\cite{chen2020revisiting,xia2022hypergraph}.\\\vspace{-0.12in}

\noindent \textbf{Intent-aware Information Aggregation}. We will describe how to incorporate intent-aware global user (item) dependencies into our GNN-based collaborative filtering framework. In our multi-intent encoder, disentangled user-item preferences are preserved in $\mathbb{E}_{P(c_{u}|u)}[c_{u}]$ and $\mathbb{E}_{P(c_{v}|v)}[c_{v}]$. In our \model, we define $K$ global intent prototypes $\{\textbf{c}_{u}^{k}\in\mathbb{R}^{d}\}_{k=1}^K$ and $\{\textbf{c}_{v}^{k}\in\mathbb{R}^{d}\}_{k=1}^K$ for user and item, respectively. With these learnable intent embeddings, we generate user and item representations by aggregating information across different $K$ intent prototypes with the global context at the $l$-th graph embedding layer, using the following design:
\begin{align}
    \textbf{r}_{i,l}^{(u)}=\mathbb{E}_{P(\textbf{c}_{u}|\textbf{e}_{i,l}^{(u)})}[\textbf{c}_{u}] = \sum_{k}^{K} \textbf{c}_{u}^{k} P(\textbf{c}_{u}^{k}|\textbf{e}_{i,l}^{(u)}), \label{eq10}\\
    \textbf{r}_{j,l}^{(v)}=\mathbb{E}_{P(\textbf{c}_{v}|\textbf{e}_{j,l}^{(v)})}[\textbf{c}_{v}] = \sum_{k}^{K} \textbf{c}_{v}^{k} P(\textbf{c}_{v}^{k}|\textbf{e}_{j,l}^{(v)}),\label{eeqq10}
\end{align}
\noindent 
The $l$-th layer-specific user and item embeddings are denoted by $\textbf{e}_{i,l}^{(u)} \in \textbf{E}_{l}^{(u)}$ and $\textbf{e}_{j,l}^{(v)} \in \textbf{E}_{l}^{(v)}$, respectively. The relevance score between user $u_i$ and each intent prototype $\textbf{c}_{u}$ is defined as $P(\textbf{c}_{u}^{k}|\textbf{e}_{i,l}^{(u)})$, which can be derived as follows:

\begin{align}\label{eq12}
    P(\textbf{c}_{u}^{k}|\textbf{e}_{i,l}^{(u)}) = \frac{\eta(\textbf{e}_{i,l-1}^{(u)\top}\textbf{c}_{u}^{k})}{\sum_{k'}^{K} \eta(\textbf{e}_{i,l-1}^{(u)\top} \textbf{c}_{u}^{k'})}, P(\textbf{c}_{v}^{k}|\textbf{e}_{j,l}^{(v)}) = \frac{\eta(\textbf{e}_{j,l-1}^{(v)\top}\textbf{c}_{v}^{k})}{\sum_{k'}^{K} \eta(\textbf{e}_{j,l-1}^{(v)\top} \textbf{c}_{v}^{k'})} \nonumber
\end{align}
\noindent Here, $\eta(\cdot) = \exp(\cdot)$. After generating the propagated message, we refine it by integrating the local collaborative filtering signals with the global disentangled collaborative relations, as follows:
\begin{align}
    \textbf{E}_{l}^{(u)} = \textbf{E}_{l-1}^{(u)} + \textbf{Z}_{l-1}^{(u)} + \textbf{R}_{l-1}^{(u)}, ~~~
    \textbf{E}_{l}^{(v)} = \textbf{E}_{l-1}^{(v)} + \textbf{Z}_{l-1}^{(v)} + \textbf{R}_{l-1}^{(v)}.
\end{align}
\noindent In this equation, $\textbf{R}_{l-1}^{(u)} \in\mathbb{R}^{I\times d}$ and $\textbf{R}_{l-1}^{(v)} \in\mathbb{R}^{J\times d}$ represent the stacked intent-aware user embeddings ($\textbf{r}_{i,l-1}^{(u)}$) and item embeddings ($\textbf{r}_{j,l-1}^{(v)}$), respectively. Incorporating intent disentanglement into the graph neural architecture enables our learned representations to effectively disentangle the latent factors driving complex user-item interaction behaviors.

\vspace{-0.1in}
\subsection{Disentangled Contrastive Learning}
Taking inspiration from recent developments in contrastive learning, we explore the potential of contrastive augmentation with intent disentanglement to address the data sparsity issue in recommender systems. Although self-supervision signals can be generated by maximizing the consistency between positive pairs among contrastive views, we argue that such augmentation is susceptible to data noise, such as misclicks. Noisy contrastive regularization may mislead the self-supervised learning process. For instance, reinforcing the model to achieve embedding agreement via node self-discrimination on noisy interaction edges may involve noisy self-supervised signals and lead to suboptimal representations.

To address this challenge, we design learnable augmenters that consider both local collaborative relations and global disentangled user (item) dependencies. By doing so, the learnable contrastive augmenters can adaptively learn disentangled SSL signals.

\subsubsection{\bf {Disentangled Data Augmentation}}
To enable the augmentation to be adaptive to each connection hop, we introduce a learnable relation matrix $\mathcal{G}^{l} \in \mathbb{R}^{I\times J}$ for each $(l)$-th GNN layer to encode the implicit relationships between users and items. Inspired by previous work on graph denoising~\cite{luo2021learning, tian2022learning}, we aim to generate a graph mask $\mathcal{M}^{l} \in \mathbb{R}^{I\times J}$, which can be used to obtain the relation matrix through element-wise multiplication: $\mathcal{G}^{l} = \mathcal{M}^{l} \odot \mathcal{A}$. \\\vspace{-0.12in}

\noindent \textbf{Learning Graph Mask}. 
Each entry $\mathcal{M}^{l}_{i, j} \in [0, 1]$ in the graph mask $\mathcal{M}^{l}$ reflects the degree to which the interaction between user $i$ and item $j$ is masked. The closer the value is to $0$, the less important the interaction is, and vice versa. In our DCCF model, we derive $\mathcal{M}^{l}_{i, j}$ based on the disentangled embeddings of user ($\textbf{r}_{i,l}^{(u)}$) and item ($\textbf{r}_{i,l}^{(v)}$) to preserve the intent-aware interaction patterns. Specifically, we use cosine similarity~\cite{chen2020iterative,tian2022learning} between node embeddings to measure the importance of interactions:
\begin{align}
    s(\textbf{r}_{i,l}^{(u)}, \mathbf{r}_{j,l}^{(v)}) = \frac{{\textbf{r}_{i,l}^{(u)}}^T{\mathbf{r}_{j,l}^{(v)}}}{\|{\textbf{r}_{i,l}^{(u)}}\|_2\|{\mathbf{r}_{j,l}^{(v)}}\|_2}.
\end{align}
The mask value is obtained by linearly transforming the range of the similarity to $[0, 1]$, using the formula: $\mathcal{M}^{l}_{i, j} = (s(\textbf{r}_{i,l}^{(u)}, \mathbf{r}_{j,l}^{(v)}) + 1) / 2$.

\noindent \textbf{Learnable Augmentation}. 
$\mathcal{A}_{i, j}$ is $0$ when there is no interaction between user $i$ and item $j$. $\mathcal{G}^{l}$ is obtained by element-wise multiplication of $\mathcal{M}^l$ and $\mathcal{A}$. only the mask values of observed interactions are calculated for computational simplicity. With the learned relation matrix, we then normalize it with the degree of the node as follows (layer index is omitted for simplicity):
\begin{align}
    \bar{\mathcal{G}}_{i, j} = \mathcal{G}_{i, j} / \sum_{j'}^{J} \mathcal{G}_{i, j}, ~~~
    \bar{\mathcal{G}}^T_{j, i} = \mathcal{G}^T_{j, i} / \sum_{i'}^{I} \mathcal{G}^T_{j, i'}.
\end{align}
To integrate our adaptive augmentation with the message passing scheme, we apply our normalized learned relation matrix $\bar{\mathcal{G}}^{l}$ over the messages of nodes for learnable propagation. With this design, we perturb the graph structure to generate contrastive learning views with adaptive augmentation. The augmentation with adaptive masking can be formally presented as follows:
\begin{align}\label{eq18}
    \textbf{H}_{l}^{(u)} = \bar{\mathcal{G}} \cdot \textbf{E}_{l}^{(v)}, ~~~
    \textbf{H}_{l}^{(v)} = \bar{\mathcal{G}}^T \cdot \textbf{E}_{l}^{(u)},
\end{align}
\noindent To generate multiple contrastive views, we consider both local collaborative signals and global disentangled relationships. In particular, we perform augmentation using two learnable mask matrices over encoded local embeddings ($\textbf{Z}_l^{(u)}$ and $\textbf{Z}_l^{(v)}$ in Eq.~\ref{eq1}), and global embeddings with intent disentanglement ($\textbf{R}_l^{(u)}$ and $\textbf{R}_l^{(v)}$ in Eq.~\ref{eq10}). We derive two mask values $\mathcal{M}_{i, j}^{l}$ separately using the following formulas: $\mathcal{M}_{i, j}^{l} = (s(\textbf{r}_{i,l}^{(u)}, \mathbf{r}_{j,l}^{(v)}) + 1) / 2$ and $\mathcal{M}_{i, j}^{',l} = (s(\textbf{z}_{i,l}^{(u)}, \mathbf{z}_{j,l}^{(v)}) + 1) / 2$. After that, our augmentation-aware message passing paradigm can be described with the following embedding refinement details:

\begin{align}
    \textbf{E}_{l}^{(u)} = \textbf{E}_{l-1}^{(u)} + \textbf{Z}_{l-1}^{(u)} + \textbf{R}_{l-1}^{(u)} + \textbf{H}_{l-1}^{\beta,(u)} + \textbf{H}_{l-1}^{\gamma,(u)}\label{eq19}
\end{align}
\noindent Here, $\textbf{H}_{l-1}^{\beta,(u)}$ and $\textbf{H}_{l-1}^{\gamma,(u)}$ represent the local- and global-level augmented representations, respectively. Similarly, item embeddings are fused in an analogous manner.

\subsubsection{\bf Contrastive Learning}
Using the above augmented representation views, we conduct contrastive learning across different view-specific embeddings of users and items. Following the approach of supervised contrastive signals in~\cite{wu2021self,xia2022hypergraph}, we generate each positive pair using the embeddings of the same user (item) from the original CF view and each of the augmented views. The encoded representations of different nodes are treated as negative pairs. Specifically, we generate three augmented views using our augmenters: i) the local collaborative view with adaptive augmentation ($\textbf{H}^{\beta,(u)}$); ii) the disentangled global collaborative view ($\textbf{R}^{(u)}$); and iii) the adaptive augmented view ($\textbf{H}^{\gamma,(u)}$). We generate contrastive self-supervision signals using InfoNCE loss as follows:

\begin{align}\label{eq25}
    \mathcal{I}(\textbf{m}, \textbf{n}) = \frac{1}{I} \sum_{i=0}^I\sum_{l=0}^L -\log \frac{\exp(s(\textbf{m}_{i,l}^{(u)}, \mathbf{n}_{i,l}^{(u)})/\tau)}{\sum_{i'=0}^I \exp(s({\textbf{m}_{i,l}^{(u)}, \mathbf{n}_{i',l}^{(u)})/\tau })},
\end{align}
\noindent Here, $\textbf{m}$ denotes the original view with vanilla embeddings ($\textbf{z} \in \textbf{Z}^{(u)}$) encoded from GNN. \textbf{n} is sampled from one of three augmented embeddings $\textbf{h}^{\beta} \in \textbf{H}^{\beta,(u)}$, $\textbf{R}^{(u)}$, and $\textbf{h}^{\gamma} \in \textbf{H}^{\gamma,(u)}$. The cosine similarity function is denoted by $s(\cdot)$. The contrastive learning loss from the user side can be formalized as follows:

\begin{align}
    \mathcal{L}_{cl}^{(u)} = \mathcal{I}(\textbf{z}, \textbf{r}) + \mathcal{I}(\textbf{z}, \textbf{h}^{\beta}) + \mathcal{I}(\textbf{z}, \textbf{h}^{\gamma})
\end{align}
By stacking $L$ graph neural layers, the layer-specific embeddings are aggregated across different layers as follows: $\textbf{E}^{(u)} = \sum_{l=0}^{L} \textbf{E}_{l}^{(u)}$ and $\textbf{E}^{(v)} = \sum_{l=0}^{L} \textbf{E}_{l}^{(v)}$. The user-item preference score is derived as:
\begin{align}
    \textbf{Y} = \textbf{E}^{(u)} (\textbf{E}^{(v)})^T, ~~~ \textbf{Y}_{i, j} = (\textbf{e}_{i}^{(u)})^T \textbf{e}_{j}^{(v)}.
\end{align}
\noindent To optimize the classical supervised recommendation task using the estimated preference score, we use the following Bayesian Personalized Ranking (BPR) loss:
\begin{align}
    \mathcal{L}_{bpr}= -\frac{1}{|\mathcal{R}|} \sum_{(i, p_s, n_s) \in \mathcal{R}} ln\sigma(\textbf{Y}_{i, p_s} - \textbf{Y}_{i, n_s}),
\end{align}
\noindent where $\mathcal{R}$ is the set of sampled interactions in each mini-batch~\cite{he2020lightgcn}. For each user $u_i$, we sample $S$ positive items (indexed by $p_s$) and $S$ negative items (indexed by $n_s$) from the training data.

Finally, we integrate the self-supervised loss with our classical recommendation loss into a multi-task learning objective as follows:
\begin{align}
    \mathcal{L} = \mathcal{L}_{bpr} + \lambda_{1}\cdot(\mathcal{L}_{cl}^{(u)} + \mathcal{L}_{cl}^{(v)}) + \lambda_{2}\cdot\|\mathbf{\Theta_1}\|_{\text{F}}^2 + \lambda_{3}\cdot\|\mathbf{\Theta_2}\|_{\text{F}}^2
\end{align}
\noindent where $\lambda_{1}$, $\lambda_{2}$ and $\lambda_{3}$ are tunable weights. $\mathbf{\Theta_1} = \{ \textbf{E}_{0}^{(u)},  \textbf{E}_{0}^{(v)}\}$ and $\mathbf{\Theta_2} = \{ \{ \textbf{c}_{u}^{k}\}_{k=1}^{K}, \{ \textbf{c}_{v}^{k}\}_{k=1}^{K} \}$ are trainable parameters in our model.

\subsection{Discussions on \model\ Model}
In this section, we present theoretical analyses of the benefits of our disentangled contrastive learning paradigm. Initially, for a specific user $u_i$, the corresponding contrastive self-supervised learning signals are incorporated with $\mathcal{I}(\textbf{r}_i^{(u)}, \textbf{z}_i^{(u)})$, where $\textbf{r}_i^{(u)}$ is the encoded embedding of $u_i$ from the augmentation with intent-aware user global dependency. The gradients of $\mathcal{I}(\textbf{r}_i^{(u)}, \textbf{z}_i^{(u)})$ with respect to the disentangled representation $\textbf{r}_i^{(u)}$ contributed by negative samples can be derived as follows:

\begin{align}
    c(i')&=\left(\frac{\textbf{r}_i^{(u)}}{\|\textbf{r}_i^{(u)}\|_2} - s(\textbf{r}_i^{(u)}, \mathbf{z}_{i'}^{(u)}) \frac{\textbf{z}_i^{(u)}}{\|\textbf{z}_i'^{(u)}\|_2} \right) \\\nonumber
    &\times \frac{\exp(s(\textbf{r}_i^{(u)}, \textbf{z}_i'^{(u)})/\tau)}{\sum_{i'} \exp(s(\textbf{r}_{i}^{(u)}, \mathbf{z}_{i'}^{(u)}/\tau)} 
\end{align}
\noindent Without loss of generality, we omit the index of graph layers. Here, $i'$ denotes the negative sample $u_i'$ for $u_i$ ($i' \neq i$ \& $1 \leq i \leq I$). The L2 norm of $c(i')$ is proportional to a special function as follows:
\begin{align}
    \|c(i')\|_2 \propto \sqrt{1-s(\textbf{r}_i^{(u)}, \mathbf{z}_{i'}^{(u)})^2}\cdot\exp(\frac{s(\textbf{r}_i^{(u)}, \mathbf{z}_{i'}^{(u)})}{\tau})
\end{align}
\noindent In the above equation, $s(\textbf{r}_i^{(u)}, \mathbf{z}_{i'}^{(u)}) \in [-1, 1]$. For hard negative samples, the corresponding embedding similarity score is close to 1, and the L2 norm of $c(i')$ increases significantly~\cite{wu2021self, xia2022hypergraph}. Similar observations can be made for the contrastive augmentations $\mathcal{I}(\textbf{z}, \textbf{h}^{\alpha})$ and $\mathcal{I}(\textbf{z}, \textbf{h}^{\beta})$ using the learnable augmenter. Thus, our disentangled contrastive learning paradigm is capable of seeking hard negative samples to enhance model optimization.

In addition, we further justify the effectiveness of our model design for capturing the implicit cross-intent dependency via the gradient propagation. Here, we discuss how the encoding process of disentangled representation $\textbf{r}_i^{(u)}$ can propagate gradients to latent intent prototypes $\{\textbf{c}_u^k\}_{k=1}^K$. Referring to Equation (\ref{eq10}) and (\ref{eq12}), we have the following partial derivative:

\begin{align}
    \frac{\partial \textbf{r}_i^{(u)}}{\partial\textbf{c}_u^{t}} = 
    \begin{bmatrix} 
    \frac{\partial(\textbf{r}_i^{(u)})_1}{\partial(\textbf{c}_u^{t})_1} & \cdots &\frac{\partial(\textbf{r}_i^{(u)})_1}{\partial(\textbf{c}_u^{t})_d}\\
    \vdots & \ddots & \vdots\\
    \frac{\partial(\textbf{r}_i^{(u)})_d}{\partial(\textbf{c}_u^{t})_1} & \cdots & \frac{\partial(\textbf{r}_i^{(u)})_d}{\partial(\textbf{c}_u^{t})_d}
    \end{bmatrix},
\end{align}
\begin{align}
    \frac{\partial(\textbf{r}_i^{(u)})_m}{\partial(\textbf{c}_u^{t})_n} = P_t\sum_{k=1}^{K}P_k[(\textbf{e}_i^{(u)})_n((\textbf{c}_u^{t})_m-(\textbf{c}_u^{k})_m) + \mathbb{I}(m=n)].
\end{align}
\noindent $P_t$ is short for $P(\textbf{c}_u^t|\textbf{e}_i^{(u)})$. As can be seen from the partial derivatives, the intent-aware representations $\textbf{r}_i^{(u)}$ propagate gradients to the latent intent prototype via the estimation of conditional probability $\sum_{k}^{K} P(y, c_{u}^{k}, c_{v}^{k}|u, v)$. During the backward propagation process, the cross-intent embedding aggregation can propagate gradients to all latent intents with the learned relevance weights. Therefore, the gradient learning enhanced by our auxiliary contrastive learning tasks is appropriately distributed to all latent intents, which facilitates the cross-intent dependency modeling and helps to capture accurate user preferences for recommendation.\\\vspace{-0.12in}

\noindent \textbf{Time Complexity Analysis}.
We analyze the time complexity of different components in our \model\ from the following aspects: \romannumeral1) The graph-based message passing procedure takes $\mathcal{O}(L\times|\mathcal{A}|\times d)$ time, where $L$ denotes the number of graph neural layers for message passing. $|\mathcal{A}|$ represents the number of edges in the graph and $d$ is the dimensionality of user/item representations. \romannumeral2) The intent-aware information aggregation component takes $\mathcal{O}(L \times (I + J) \times K \times d)$ time complexity, where $K$ denotes the number of latent intents. \romannumeral3) Due to local- and global-based adaptive augmentation, it takes $\mathcal{O}(2 \times L \times |\mathcal{A}| \times d)$ time complexity to generate two augmented views for self-supervision. \romannumeral4) To calculate the contrastive learning objective, the cost is $\mathcal{O}(L \times B \times (I + J) \times d)$, where $B$ is the number of users/items included in a single mini-batch.

\section{Evaluation}
\label{sec:eval}

In this section, we perform experiments to evaluate our \model\ on different datasets by answering the following research questions:
\begin{itemize}[leftmargin=*]
\item \textbf{RQ1}: Does our proposed \model\ outperform various recommendation solutions under different experimental settings?
\item \textbf{RQ2}: Do the designed key components benefit the representation learning of our \model\ in achieving performance improvement?
\item \textbf{RQ3}: Is our proposed model effective in alleviating the data sparsity issues with our disentangled self-supervised signals?
\item \textbf{RQ4}: What is the impact of the number of latent intents?
\item \textbf{RQ5}: How does our \model\ perform \wrt\ training efficiency?
\end{itemize}

\begin{table}
    \centering
    \small
    \caption{Statistics of the experimental datasets.}
    \vspace{-0.15in}
    \begin{tabular}{ccccc}
        \toprule
        Dataset & \#Users & \#Items & \#Interactions & Density\\
        \midrule
        Gowalla & 50,821 & 57,440 & 1,172,425 & 4.0$e^{-4}$\\
        Amazon-book & 78,578 & 77,801 & 2,240,156 & 3.7$e^{-4}$\\
        Tmall & 47,939 & 41,390 & 2,357,450 & 1.2$e^{-3}$\\
        \bottomrule
    \end{tabular}
    \vspace{-0.05in}
    \label{tab:data statistics}
\end{table}

\subsection{Experimental Settings}
\subsubsection{\bf Datasets} We evaluate our model performance on public datasets: \textbf{Gowalla:} This dataset is collected from the Gowalla platform to record check-in relations between users and different locations based on mobility traces. \textbf{Amazon-book:} This dataset includes rating behaviors of users over products with book category on Amazon. \textbf{Tmall:} It contains customer purchase behaviors from the online retailer Tmall. Table \ref{tab:data statistics} summarizes the dataset statistics.

\begin{table*}[!tb]
  \centering
  \caption{Recommendation performance of all compared methods on different datasets in terms of Recall and NDCG.}
  \vspace{-0.1in}
  \resizebox{\textwidth}{!}{
    \begin{tabular}{c|cccc|cccc|cccc}
      \toprule
            Data & \multicolumn{4}{c|}{Gowalla} & \multicolumn{4}{c|}{Amazon-book} & \multicolumn{4}{c}{Tmall} \\
    \midrule
            Metrics & Recall@20 & Recall@40 & NDCG@20 & NDCG@40 & Recall@20 & Recall@40 & NDCG@20 & NDCG@40 & Recall@20 & Recall@40 & NDCG@20 & NDCG@40\\
      \midrule
      NCF& 0.1247& 0.1910& 0.0659& 0.0832& 0.0468& 0.0771& 0.0336& 0.0438& 0.0383& 0.0647& 0.0252& 0.0344 \\
      \midrule
      AutoR& 0.1409& 0.2142& 0.0716& 0.0905& 0.0546& 0.0914& 0.0354& 0.0482& 0.0336& 0.0611& 0.0203& 0.0295 \\
      \midrule
      NGCF& 0.1413& 0.2072& 0.0813& 0.0987& 0.0532& 0.0866& 0.0388& 0.0501& 0.0420& 0.0751& 0.0250& 0.0365 \\
      LightGCN& 0.1799& 0.2577& 0.1053& 0.1255& 0.0732& 0.1148& 0.0544& 0.0681& 0.0555& 0.0895& 0.0381& 0.0499 \\
      \midrule
      DisenGCN& 0.1379& 0.2003& 0.0798& 0.0961& 0.0481& 0.0776& 0.0353& 0.0451& 0.0422& 0.0688& 0.0285& 0.0377\\
      DisenHAN& 0.1437& 0.2079& 0.0829& 0.0997& 0.0542& 0.0865& 0.0407& 0.0513& 0.0416& 0.0682& 0.0283& 0.0376\\
      CDR& 0.1364& 0.1943& 0.0812& 0.0963& 0.0564& 0.0887& 0.0419& 0.0526& 0.0520& 0.0833& 0.0356& 0.0465\\
      DGCF& 0.1784& 0.2515& 0.1069& 0.1259& 0.0688& 0.1073& 0.0513& 0.0640& 0.0544& 0.0867& 0.0372& 0.0484\\
      DGCL& 0.1793& 0.2483& 0.1067& 0.1247& 0.0677& 0.1057& 0.0506& 0.0631& 0.0526& 0.0845& 0.0359& 0.0469\\
      \midrule
      SLRec& 0.1529& 0.2200& 0.0926& 0.1102& 0.0544& 0.0879& 0.0374& 0.0490& 0.0549& 0.0888& 0.0375& 0.0492\\
      SGL-ED& 0.1809& 0.2559& 0.1067& 0.1262& 0.0774& 0.1204& 0.0578& 0.0719& 0.0574& 0.0919& 0.0393& 0.0513\\
      SGL-ND& 0.1814& 0.2589& 0.1065& 0.1267& 0.0722& 0.1121& 0.0542& 0.0674& 0.0553& 0.0885& 0.0379& 0.0494\\
      HCCF& 0.1818& 0.2601& 0.1061& 0.1265& 0.0824& 0.1282& 0.0625& 0.0776& 0.0623& 0.0986& 0.0425& 0.0552\\
      LightGCL& 0.1825& 0.2601& 0.1077& 0.1280& 0.0836& 0.1280& 0.0643& 0.0790& 0.0632& 0.0971& 0.0444& 0.0562\\
      \midrule
      {\model} & \textbf{0.1876} & \textbf{0.2644} & \textbf{0.1123} & \textbf{0.1323} & \textbf{0.0889} & \textbf{0.1343} & \textbf{0.0680} & \textbf{0.0829} & \textbf{0.0668} & \textbf{0.1042} & \textbf{0.0469} & \textbf{0.0598} \\
      \midrule
      p-val.  & $8.9e^{-6}$ & $1.3e^{-3}$ & $2.6e^{-6}$ & $8.1e^{-6}$ & $8.6e^{-7}$ & $2.2e^{-6}$ & $8.6e^{-6}$ & $2.2e^{-6}$ & $2.6e^{-7}$ & $1.4e^{-7}$ & $8.6e^{-7}$ & $1.8e^{-7}$   \\
      \bottomrule
      \end{tabular}}
  \label{tab:overall comparison}%
\end{table*}%

\subsubsection{\bf Evaluation Protocols and Metrics}
To alleviate the bias of negative item instance sampling, we follow the all-rank protocol~\cite{he2020lightgcn,wu2021self,wang2020disenhan} over all items to measure the accuracy of our recommendation results. We use two widely adopted ranking-based metrics to evaluate the performance of all methods, namely {\it Recall@N} and {\it NDCG (Normalized Discounted Cumulative Gain)@N}.

\subsubsection{\bf Baseline Methods}
We include five groups of baseline methods for comprehensive comparison, as detailed below.\\\vspace{-0.12in}

\noindent \textbf{(i) Factorization-based Method.}
\begin{itemize}[leftmargin=*]

\item \textbf{NCF}~\cite{he2017neural}. This method replaces the inner product in MF with a multi-layer perceptron to estimate user-item interactions. For comparison, we include the NeuMF version.
\end{itemize}

\noindent \textbf{(ii) Autoencoder-based Method.}
\begin{itemize}[leftmargin=*]
\item \textbf{AutoR}~\cite{sedhain2015autorec}. It reconstructs user-item interactions based on the autoencoder to obtain user preference for non-interacted items.
\end{itemize}

\noindent \textbf{(iii) Recommendation with Graph Neural Network.}
\begin{itemize}[leftmargin=*]

\item \textbf{NGCF}~\cite{wang2019neural}. This method designs the propagation rule to inject collaborative signals into the embedding process of recommendation, which is beneficial for capturing higher-order connectivity.

\item \textbf{LightGCN}~\cite{he2020lightgcn}. This method simplifies the message passing rule of GCN by linearly propagate user/item embeddings on the interaction graph for collaborative filtering.

\end{itemize}

\noindent \textbf{(iv) Disentangled Multi-Intent Recommender Systems.}
\begin{itemize}[leftmargin=*]
\item \textbf{DisenGCN}~\cite{ma2019disentangled}. This method proposes a neighborhood routing mechanism to learn disentangled node representation. The dot-product is used to predict the interaction likelihood.

\item \textbf{DisenHAN}~\cite{wang2020disenhan}. It disentangles user/item representations into different aspects ({\it i.e.}, latent intents) and then aggregates information from various aspects with attention for recommendation.

\item \textbf{CDR}~\cite{chen2021curriculumnips}. This method utilizes a user's noisy multi-feedback to mine user intentions and improves the training process through curriculum learning. We implement it with implicit feedback.

\item \textbf{DGCF}~\cite{wang2020disentangled}. This method generates the intent-aware graph by modeling a distribution over intents for each interaction and thus learns disentangled representations.

\item \textbf{DGCL}~\cite{li2021disentangled}. This method proposes a factor-wise discrimination objective to learn disentangled representations. We implement it to learn disentangled representations of nodes and make user-item interaction prediction using inner products.
\end{itemize}

\noindent \textbf{(v) Self-Supervised Learning for Recommendation.}
\begin{itemize}[leftmargin=*]

\item \textbf{SLRec}~\cite{yao2021self}. This method proposes a multi-task self-supervised learning framework to address the label sparsity problem in large-scale item recommender system.

\item \textbf{SGL-ED/ND}~\cite{wu2021self}. This method reinforces user/item representation learning with GNNs by applying an auxiliary self-supervised contrastive learning task through data augmentation, namely edge drop (ED) or node drop (ND).

\item \textbf{HCCF}~\cite{xia2022hypergraph}. It jointly captures local and global collaborative relations under a hypergraph neural network, and designs cross-view contrastive learning for augmentation.

\item \textbf{LightGCL}~\cite{anonymous2023simple}. It is a lightweight graph contrastive learning framework by leveraging singular value decomposition to generate augmented view for embedding contrasting.

\end{itemize}

\vspace{-0.1in}
\subsubsection{\bf Hyperparameter Settings.}
We implement our {\model} using PyTorch and use Adam~\cite{KingmaB14adam} as optimizer with learning rate $1e^{-3}$. The number of latent intent prototypes $K$ is selected from the range of $\{ 32, 64, 128, 256 \}$ with $K=128$ by default. $\lambda_1$, $\lambda_2$ and $\lambda_3$ are tuned from the range of $[0.001, 0.025, 0.1, 0.2]$, $[2.5e^{-5}, 5e^{-4}, 5e^{-3}]$, $[2.5e^{-5}, 5e^{-4}, 5e^{-3}]$, respectively. To evaluate baseline performance with fair settings, latent embedding dimensionality $d$ and batchsize is set as 32 and 10240 for all compared methods. For graph-based models, the number of propagation layers is chosen from \{1,2,3\}. Detailed model implementation of our \model\ can be found in our released source code in the Abstract Section.

\begin{table}[t]
    \centering
    \caption{Ablation study on key components of {\model} (measured by $Recall@20$ and $NDCG@20$) on different datasets.}
    \footnotesize
    \begin{tabular}{c|c|cc|cc|cc}
        \toprule
        \multirow{2}{*}{Category} & Data & \multicolumn{2}{c|}{Gowalla} & \multicolumn{2}{c|}{Amazon-book} & \multicolumn{2}{c}{Tmall}\\
        \cmidrule{2-8}
        & Variants & Recall & NDCG & Recall & NDCG & Recall & NDCG\\
        \midrule
        \multirow{1}{*}{DME} 
        &-Disen & 0.1637 & 0.0975 & 0.0772 & 0.0580 & 0.0629 & 0.0437\\
        \midrule
        \multirow{2}{*}{PAM} 
        &-LocalR & 0.1719 & 0.1015 & 0.0786 & 0.0593 & 0.0638 & 0.0446\\
        &-DisenR & 0.1718 & 0.1016 & 0.0793 & 0.0597 & 0.0640 & 0.0447\\
        \midrule
        \multirow{2}{*}{SSL} 
        &-DisenG & 0.1763 & 0.1053 & 0.0829 & 0.0635 & 0.0644 & 0.0449\\
        &-AllAda & 0.1845 & 0.1096 & 0.0833 & 0.0632 & 0.0651 & 0.0452\\
        \midrule
        \multicolumn{2}{c|}{\emph{\model}} & 0.1876 & 0.1123 & 0.0889 & 0.0680 & 0.0668 & 0.0469\\
        \hline
    \end{tabular}
    \label{tab:module_ablation}
\end{table}

\subsection{Performance Comparison (RQ1)}

Table \ref{tab:overall comparison} shows the performance comparison of different methods on all datasets. To validate the significant performance improvement achieved by our {\model} model, the p-value is provided. From evaluation results, we summarize the following observations:

\begin{itemize}[leftmargin=*]

\item {\model} consistently outperforms all baselines on all three datasets. Through disentangled contrastive learning, {\model} improves the generalization and robustness of recommenders by offering more informative representations. We attribute the significant performance gain of {\model} to two key aspects: (\romannumeral1) {\model} effectively alleviates the data sparsity issue by distilling disentangled self-supervised signals as supplementary training tasks. (\romannumeral2) Our proposed parameterized graph mask generator is beneficial for achieving adaptive self-supervision against data noise redundancy, which further improves the representation robustness.

\item Although data augmentation techniques are also proposed in current SSL-based methods (\eg, SGL, HCCF), our \model\ still outperforms them by a large margin. This is because simply learning augmented representations at coarse-grained level cannot disentangle latent intention factors behind user-item interactions. In addition, we notice that most SSL-based methods perform better than conventional GNN-based approaches (\eg, LightGCN, NGCF), which suggests the positive effects of SSL brings to GNN-based CF models. With our disentangled adaptive augmentation, {\model} still pushes that boundary forward, achieving state-of-the-art performance across all datasets.\\

\item The performance improvement of {\model} over other disentangled recommender systems (\eg, DGCF, DisenGCN, CDR) verifies that our approach is not limited to the label shortage issue. The integration of disentangled multi-intent encoding and contrastive learning results in better performance. Existing disentangled learning solutions struggle to generate informative embeddings in the face of insufficient training labels due to the overfitting effect. Although DGCL attempts to use contrastive learning to encode latent factors into augmented representations, its non-adaptive contrastive view generation makes it easily influenced by noise perturbation.

\end{itemize}

\begin{figure}[t]
    \centering
    \subfigure[Performance \textit{w.r.t.} different item groups]{
        \includegraphics[width=1.0\columnwidth]{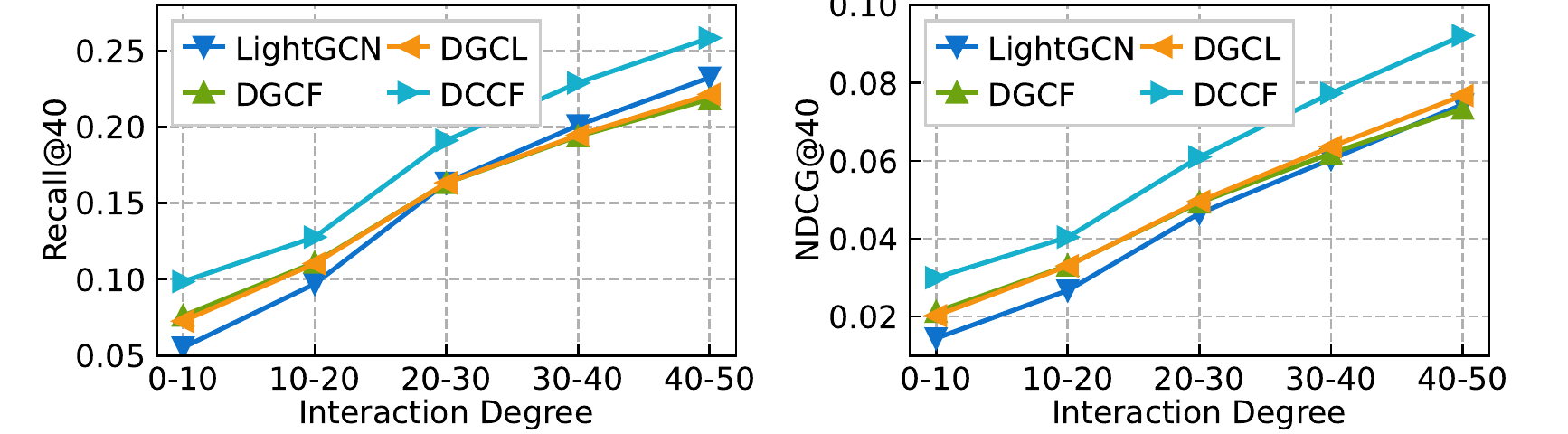}\ 
        \vspace{-0.15in}
    }
    \subfigure[Performance \textit{w.r.t.} different user groups]{
        \includegraphics[width=1.0\columnwidth]{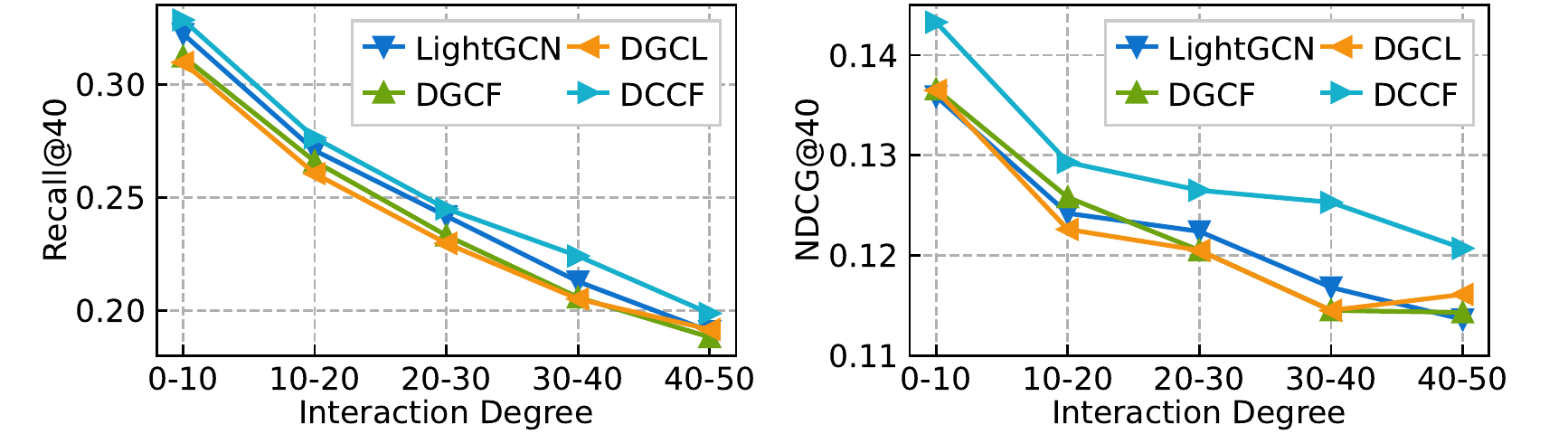}
    }
    \vspace{-0.15in}
    \caption{Performance comparison \textit{w.r.t.} data sparsity over different user/item groups on Gowalla data.}
    \vspace{-0.15in}
    \label{fig:sparse}
\end{figure}

\subsection{Ablation Study (RQ2)}
In this section, to verify the effectiveness of each component, we conduct an ablation study to examine the component-specific benefits of our {\model} framework from three perspectives: (\romannumeral1) Disentangled Multi-intent Encoding (DME); (\romannumeral2) Parameterized Adaptive Masking (PAM); (\romannumeral3) Self-supervised Learning (SSL). The performance results are reported in Table~\ref{tab:module_ablation}, and the variant details and impact study are presented as follows:
\begin{itemize}[leftmargin=*]
\item \textbf{Disentangled Multi-intent Encoding (DME)}. We generate the ablation model ({\it -Disen}) by removing the disentangled multi-intent encoding module. The performance gap between \model\ and {\it -Disen} indicates the contribution of multi-intent representation encoding to the overall performance. \\\vspace{-0.12in}

\item \textbf{Parameterized Adaptive Masking (PAM)}. To investigate the effect of our parameterized adaptive masking, we create two variants: (\romannumeral1) {\it -LocalR} which removes implicit user-item relation learning based on local relation embeddings; and (\romannumeral2) {\it -DisenR}, which removes the intent-based graph structure learning process. The results show that both variants lead to a performance degradation, indicating the necessity of adaptive self-supervised signal distillation for contrastive augmentation. \\\vspace{-0.12in}

\item \textbf{Self-Supervised Learning (SSL)}. We also examine the influence of our disentangled contrastive learning on performance by adjusting the incorporated self-supervised optimization objectives. Specifically, we creat two variants by removing agreements between the original graph representations with auxiliary augmented views: (\romannumeral1) disentangled global collaborative view ({\it -DisenG}) and (\romannumeral2) all augmented views with adaptive masking ({\it -AllAda}). Our results show that {\model} achieves the best performance compared to these variants, further emphasizing the benefits of integrating auxiliary self-supervised learning signals from the global view of intent-aware collaborative relationships for adaptive data augmentation.

\end{itemize}

\begin{table}[t]
    \setlength{\tabcolsep}{2mm}
    \centering
    \caption{The embedding smoothness on Amazon-book and Tmall data measured by MAD metric (the smaller the MAD indicates more obvious the over-smoothing phenomenon).}
    \vspace{-0.05in}
    \footnotesize
    \begin{tabular}{c|ccccc}
        \toprule
        \multirow{2}{*}{\shortstack{Embedding \\ Type}} & DCCF & DCCF-CL & DGCL & DisenGCN & LightGCN\\
        \cmidrule{2-6}
         &\multicolumn{5}{c}{Amazon-book}\\
        \midrule
        {\it User} & \textbf{0.999} & 0.902 & 0.980 & 0.961 & 0.984\\
        {\it Item} & \textbf{0.990} & 0.961 & 0.989 & 0.986 & 0.944\\
        \midrule
        & \multicolumn{5}{c}{Tmall}\\
        \midrule
        {\it User} & \textbf{0.999} & 0.800 & 0.897 & 0.876 & 0.910\\
        {\it Item} & \textbf{0.998} & 0.873 & 0.920 & 0.992 & 0.927\\
        \hline
    \end{tabular}
    \label{tab:mad}
\end{table}

\subsection{In-Depth Analysis of \model\ (RQ3 \& RQ4)}

\subsubsection{\textbf{Performance {\wrt} Data Sparsity}} We further verify if DCCF is robust to data sparsity issue. To do this, we divide users and items into different groups based on the number of their interactions, and separately measured recommendation accuracy for each group. From the results in Figure \ref{fig:sparse}, we make two main observations: (\romannumeral1) {\model} consistently outperforms several representative baselines (\ie, LightGCN, DGCL, DGCF) by providing better recommendation results for both inactive and active users. This indicates the benefits of our generated self-supervised signals in alleviating sparse data issues. While DGCL conducts factor-wise alignment with contrastive learning, the interaction noise and bias can still impair the disentangled representation learning for latent factors.(\romannumeral2) We notice that the performance gap between {\model} and the compared methods is still apparent on low-degree items. This is because the baseline DGCF only focuses on splitting the user representation into multiple intent-aware embeddings, which can easily lead to recommending high-degree items and neglect the long-tail items. In contrast, our {\model} enhances the interaction modeling on long-tail items through effective self-supervised information.

\subsubsection{\textbf{Impact of the Number of Intent Prototypes.}} 
\begin{figure}[t]
    \centering
    \includegraphics[width=1.05\columnwidth]{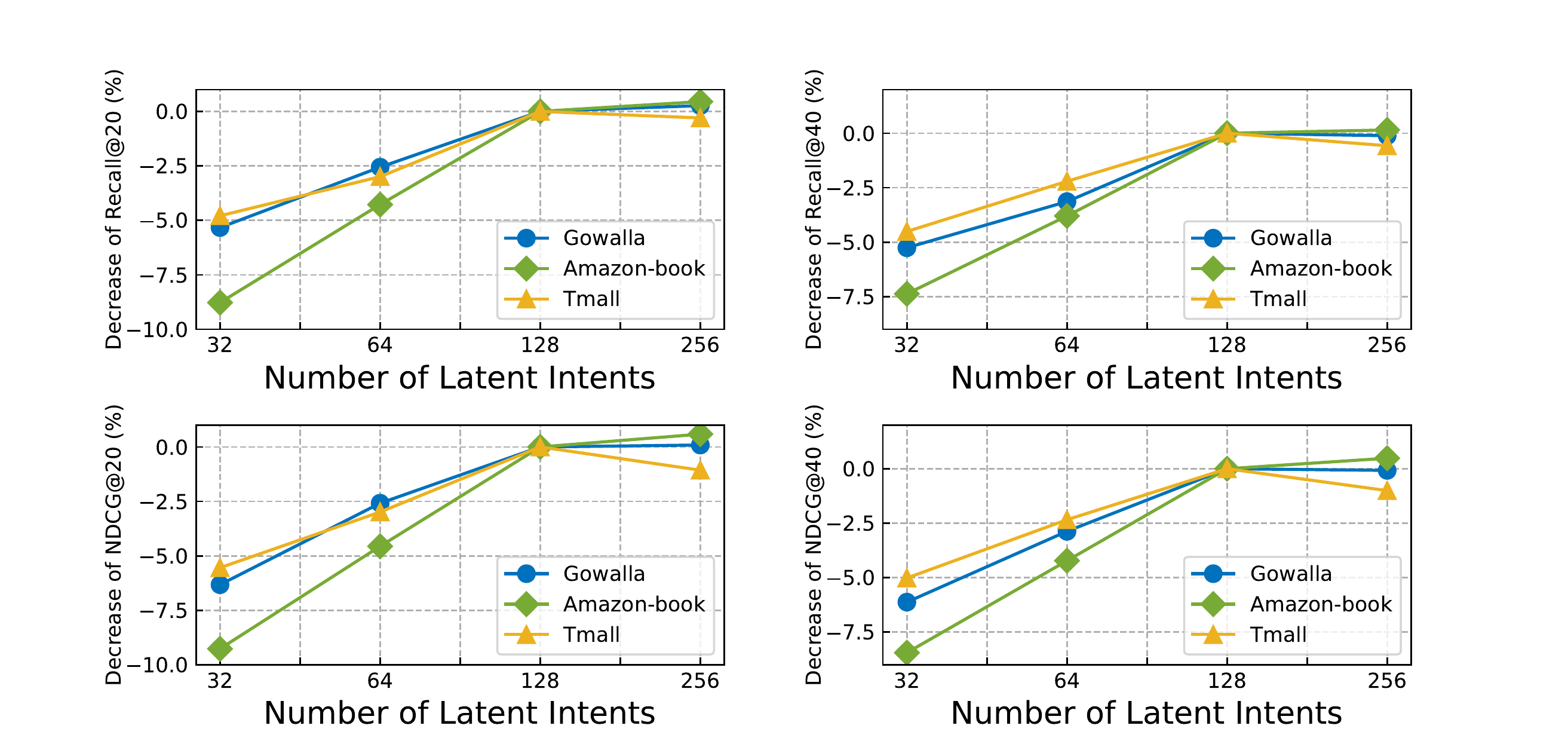}
    \caption{Performance \wrt\ the number of latent intents.}
    \vspace{-0.1in}
    \label{fig:intents}
    \vspace{-0.1in}
\end{figure}

\begin{figure}[t]
    \centering
    \subfigure[User intent prototypes]{
        \includegraphics[width=0.97\columnwidth]{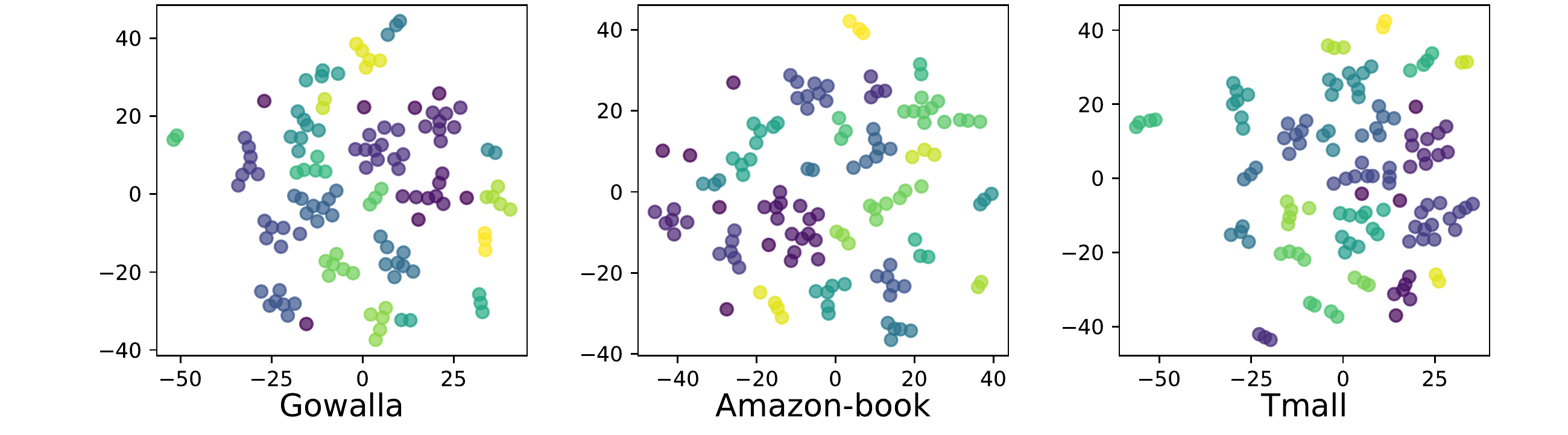}\ 
    }
    \vspace{-0.1in}
    \subfigure[Item intent prototypes]{
        \includegraphics[width=0.97\columnwidth]{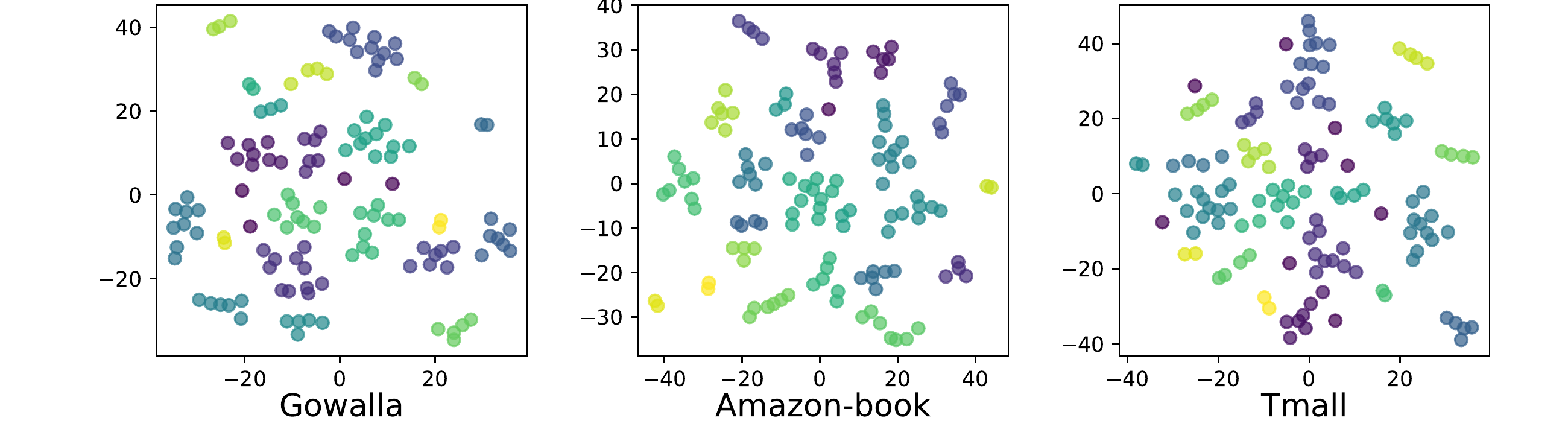}
    }
    \caption{Distribution of latent intent prototypes.}
    \label{fig:distribution}
\end{figure}

To investigate the impact of the number of latent intents on model performance, we select this parameter from the range $\{32, 64, 128, 256\}$ and re-train the model. The results are shown in Figure~\ref{fig:intents}. It is clear that as the number of intents increases, the performance of the model also improves. However, when the number of intents increases from 128 to 256, the performance improvement is limited, and even degrades on the Tmall dataset. To further understand this phenomenon, we transform the intent prototypes into 2D space for visualization using t-SNE~\cite{van2008visualizing} and then clustered them. As shown in Figure~\ref{fig:distribution}, when the number of intents is 128, some latent intents have begun to cluster together. Further increasing the number of intents causes intent redundancy with too fine-grained latent factor granularity and introduces noise into learning representations.

\subsubsection{\textbf{Robustness of DCCF in Alleviating Over-Smoothing}} 
To validate the effectiveness of {\model} in alleviating over-smoothing, we calculate the Mean Average Distance (MAD)~\cite{chen2020measuring, xia2022hypergraph} over encoded user/item embeddings of {\model} and the variant {\model}-CL, which disables the cross-view contrastive learning module. We also calculate the MAD of several representative baseline methods ({\it i.e.}, DGCL, DisenGCN, LightGCN) for comparison. Note that all the embeddings were normalized before calculating MAD for fair comparison. The results are shown in Table~\ref{tab:mad}. We notice that by removing the SSL objective, the over-smoothing phenomenon becomes more pronounced, which suggests the effectiveness of our contrastive learning component in addressing the over-smoothing problem. Moreover, all the baselines have lower MAD than our {\model}, indicating that {\model} is capable of alleviating the over-smoothing issue in the widely-adopted GNN architecture. Our disentangled contrastive learning approach achieves better representation uniformity in recommendation compared to the baselines.

\begin{table}[h]
    \setlength{\tabcolsep}{2mm}
    \centering
    \caption{Computational cost evaluation in terms of per-epoch training time (seconds) on Gowalla, Amazon, and Tmall data.}
    \vspace{-0.1in}
    \footnotesize
    \begin{tabular}{c|cccc|c}
        \toprule
        Model & DisenGCN & DGCF & DisenHAN & DGCL & Ours\\
        \midrule
        Gowalla& 19.1s & 25.1s & 16.8s & 9.3s & 12.4s\\
        Amazon-book & 42.2s & 49.6s & 30.6s & 12.4s & 18.9s\\
        Tmall & 43.5s & 51.6s & 29.8s & 12.0s & 18.8s\\
        \hline
    \end{tabular}
    \label{tab:efficency}
\end{table}

\subsection{Model Training Efficiency Study (RQ5)}
In this section, we investigate the model efficiency of our {\model} in terms of training computational cost on all datasets. The experiments were conducted on a server with system configurations of an Intel Xeon Gold 6330 CPU, NVIDIA RTX 3090. As shown in Table~\ref{tab:efficency}, we compare our {\model} with disentangled recommender systems (\eg, DGCF and DisenHAN) and found that our {\model} achieves comparable training efficiency in all cases. Specifically, while DGCF splits the user embedding into intent-aware vectors to reduce embedding size, the heavy cost of DGCF stems from the recursively routing mechanism for information propagation. It requires extra time to process multiple iterations to obtain intent-relevant weights. In DisenHAN, the time-consuming graph attention network brings high cost due to the need for computing the attention weights.

\begin{figure}[t]
    \centering
    \includegraphics[width=0.92\columnwidth]{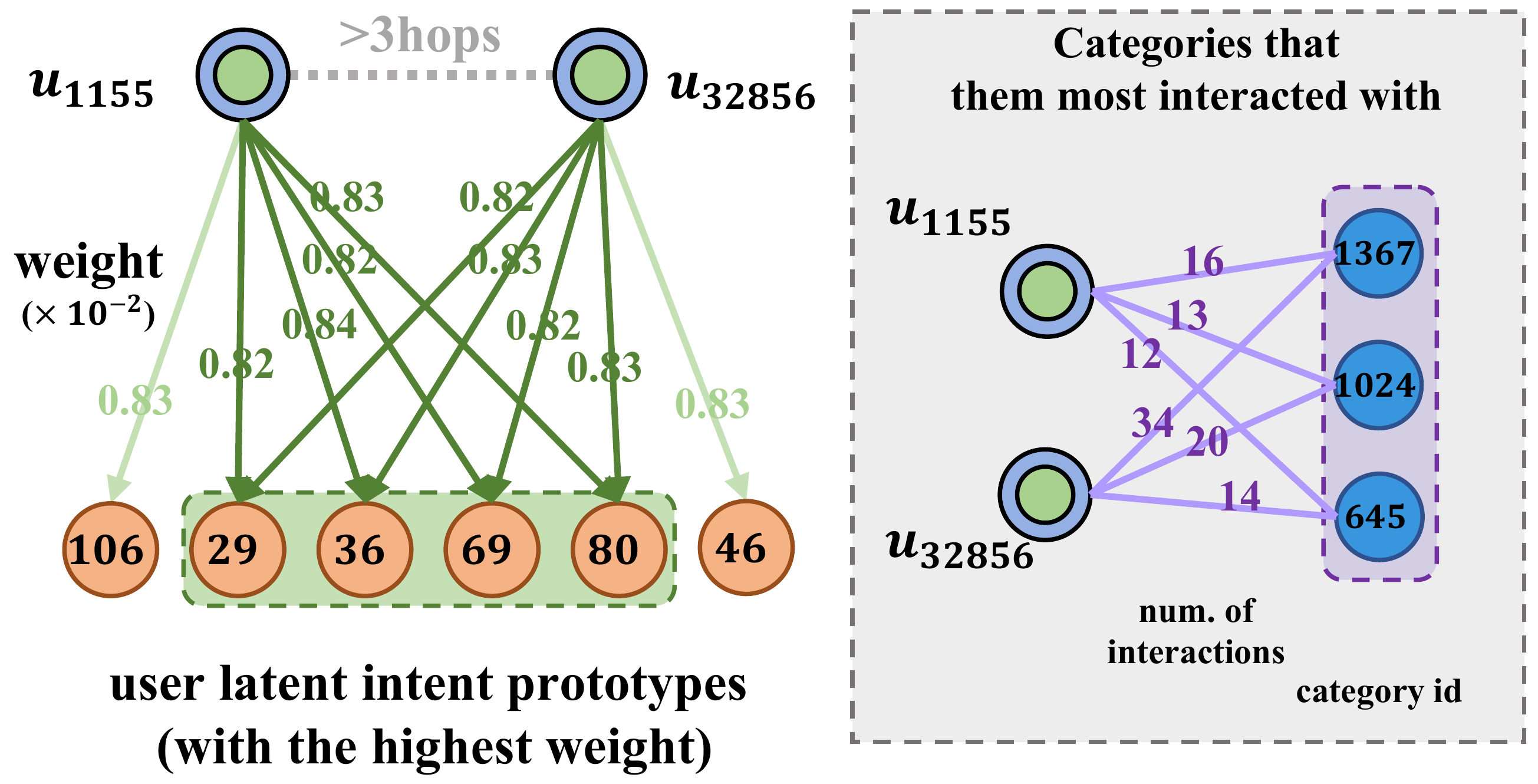}
    \caption{Case study of intent-aware global user relations. Non-locally connected users ($u_{1155}$ and $u_{32856}$) can be identified with similar user preference (large item category overlap) via our learned disentangled representations.}
    \label{fig:case}
\end{figure}

\subsection{Case Study}

\textbf{Global Intent-aware Semantic Dependency}.
In this section, we examine the potential ability of our {\model} in capturing the global intent-aware semantic dependencies among users. To achieve this goal, we showe some concrete examples in Figure~\ref{fig:case} to visualize the intent-aware user preferences learned by our {\model}. We observe that $u_{1155}$ and $u_{32856}$ share very similar intent-aware preferences, as shown with intent prototype-specific user weights, despite not being locally connected on the interaction graph. After investigating their interaction patterns, we observe a significant overlap between the categories (categories $29, 36,$ and $69$) of the items they interacted with, indicating the high semantic relatedness of their interaction behaviors. Therefore, in addition to local collaborative relations, the global intent-aware user dependencies can be preserved with our encoded disentangled user representations. \\\vspace{-0.12in}

\begin{figure}[t]
    \centering
    \includegraphics[width=0.92\columnwidth]{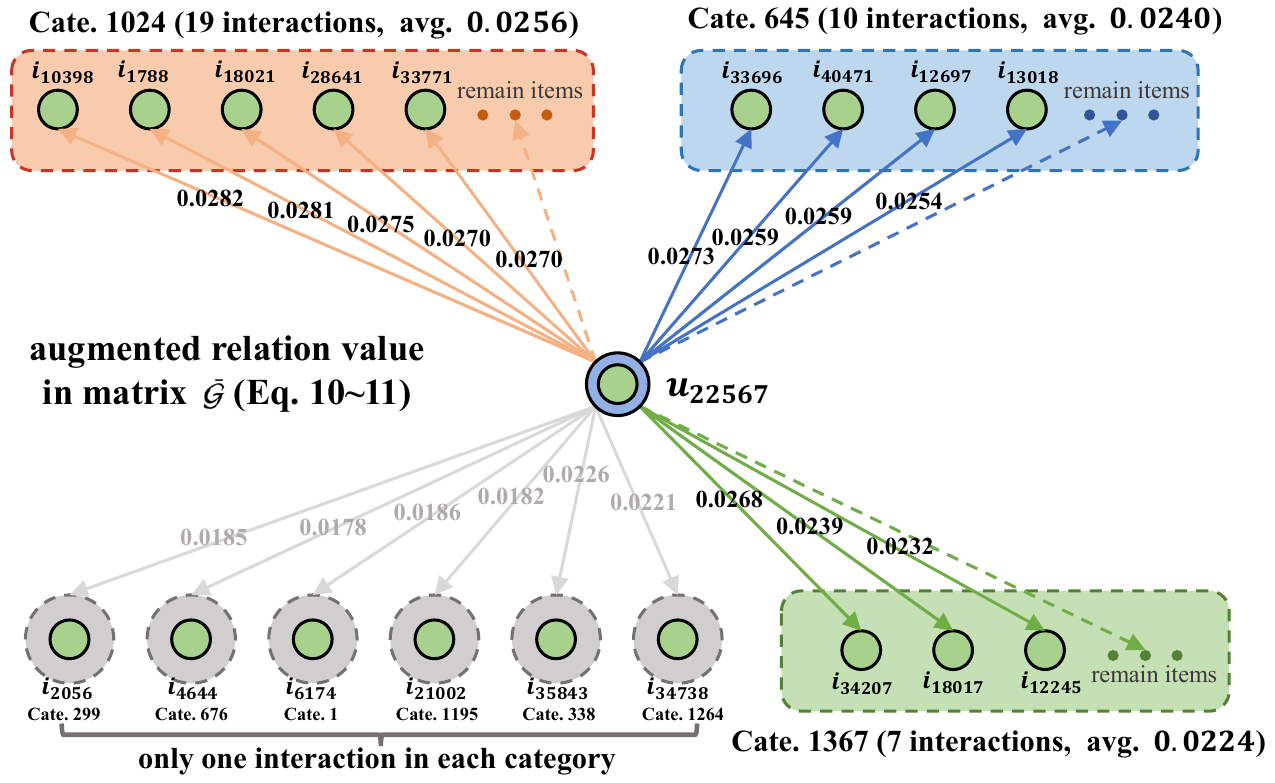}
    \vspace{-0.1in}
    \caption{Case study of intent-aware adaptive augmentation over the user-item relation matrix. User interacted items are grouped in terms of their categories. The value of the learned user-item connectivity weight is consistent with the user preference degree, \ie, the higher user-item weight encoded by \model\ indicates stronger user preference.}
    \label{fig:case2}
    \vspace{-0.1in}
\end{figure}

\noindent \textbf{Intent-aware Adaptive Augmentation}
We further analyze the rationality of our intent-aware adaptive augmentation over user-item relations. As shown in Figure~\ref{fig:case2}, we grouped the interacted items of user $u_{22567}$ based on categories (\eg, category 1024, 645). After performing adaptive augmentation over the user-item relation matrix, the implicit dependency weight between each user-item pair was learned through our contrastive intent disentanglement. The value of the learned user-item connectivity weight determines the user's preference degree over this item. We notice that a higher user-item relation weight (\eg, 0.0282 or 0.0273) indicates a stronger interaction preference over category-specific items, which is consistent with the observation of the category-specific interaction frequency of $u_{22567}$. For example, the highest item correlation weight (\ie, 0.0282) is generated from the categorical items that $u_{22567}$ interacted with the most. This observation further demonstrates the effectiveness of our disentangled contrastive augmentation, which is easily adaptable to different user-item interaction environments.
\section{Conclusion}
\label{sec:conclusoin}

This paper proposes a disentangled contrastive learning method for recommendation that explores latent factors underlying implicit intents for interactions. We introduce a graph structure learning layer that enables adaptive interaction augmentation based on learned disentangled user (item) intent-aware dependencies. Along the augmented intent-aware graph structures, we propose an intent-aware contrastive learning scheme that brings the benefits of disentangled self-supervision signals. Our extensive experiments validate the effectiveness of our proposed model on different recommendation datasets. For future work, one potential extension is to integrate disentangled representation learning with causal analysis to address the bias issues of noisy interaction data. Additionally, by considering the diverse nature of user characteristics, personalized augmentation may further enhance the power of contrastive learning for customized graph perturbing operations in recommenders. By tailoring the augmentation operations to the specific user characteristics, we may better capture the individual preferences.

\clearpage
\bibliographystyle{ACM-Reference-Format}
\balance
\bibliography{sample-base}

\clearpage

\end{document}